\documentclass[preprint,prb,aps,showpacs]{revtex4}
\usepackage{graphics}

\begin{document}
\title{Hole concentration and phonon renormalization in Ca-doped YBa$_{2}$Cu$_{3}$O$_{y}$ (6.76$\le y \le$ 7.00)}
\author{K. C. Hewitt}
\affiliation{Dalhousie University, Department of Physics, Halifax,
Nova Scotia, Canada B3H 3J5}
\email{Kevin.Hewitt@Dal.ca}\homepage{http://fizz.phys.dal.ca/~hewitt}
\author{X. K Chen, C. Roch, J. Chrzanowski, J. C. Irwin}
\affiliation{Simon Fraser University, Department of Physics,
Burnaby, British Columbia, Canada, V5A 1S6}
\author{E. H. Altendorf}
\affiliation{Microvision Inc., Bothell, Washington , USA 98011}
\author{R. Liang, D. Bonn, W. N. Hardy}
\affiliation{University of British Columbia, Department of
Physics, Vancouver, British Columbia, Canada V6T 1Z1}

\begin{abstract}
In order to access the overdoped regime of the
YBa$_2$Cu$_3$O$_{y}$ phase diagram, 2 \% Ca is substituted for Y
in YBa$_2$Cu$_3$O$_{y}$ ( y = 7.00,6.93,6.88,6.76).  Raman
scattering studies have been carried out on these four single
crystals. Measurements of the superconductivity-induced
renormalization in frequency ($\Delta \omega$) and linewidth
($\Delta 2\gamma$) of the 340 cm$^{-1}$ B$_{1g}$ phonon
demonstrate that the magnitude of the renormalization is directly
related to the hole concentration ($p$), and not simply the oxygen
content. The changes in $\Delta \omega$ with $p$ imply that the
superconducting gap ($\Delta_{max}$) decreases monotonically with
increasing hole concentration in the overdoped regime, and $\Delta
\omega$ falls to zero in the underdoped regime. The linewidth
renormalization $\Delta 2\gamma$ is negative in the underdoped
regime, crossing over at optimal doping to a positive value in the
overdoped state.
\end{abstract}

\date{\today}

\maketitle

\section{Introduction}

Certain phonons in cuprate superconductors exhibit anomalous
changes in frequency and linewidth when they are cooled through
the superconducting transition temperature. Such phonon anomalies
were first discovered in High Temperature Superconductors (HTS) by
Macfarlane {\it et al} \cite{mac87} in YBa$_2$Cu$_3$O$_{7-\delta}$
(Y123). Zeyher and Zwicknagl \cite{zey88,zey90} showed that the
presence of such anomalies could be attributed to electron-phonon
interactions and the changes in the density of electronic states
that occur on the opening of the superconducting gap. Friedl {\it
et al} \cite{fri90} used this approach to obtain an estimate for
the superconducting gap in Y123. Nicol, Jiang and Carbotte
\cite{nic93} extended the theoretical approach of Zeyher and
Zwicknagl to include the effect of a pairing interaction of d-wave
symmetry on the phonon self-energy. Phonon anomalies have been
investigated extensively, using Raman scattering, in several
cuprates - YBa$_2$Cu$_3$O$_{7-\delta}$
\cite{alt91,alt91a,alt92,alt93,hau96,pan98a},
YBa$_2$(Cu$_{1-x}M_x$)$_4$O$_{8}$ (Y124) for M=Zi,Mn \cite{wat98},
Bi$_2$Sr$_2$CaCu$_2$O$_{8+\delta}$(Bi2212) \cite{lea93,mar97},
NdBa$_2$Cu$_3$O$_{7-\delta }$ (Nd123)\cite{mis97},
HgBa$_2$Ca$_2$Cu$_3$O$_{8+\delta }$ (Hg1223)\cite{zho97},
HgBa$_2$Ca$_3$Cu$_4$O$_{10+\delta }$ (Hg1234)\cite{hadj98},
HgBa$_2$CuO$_{4+\delta }$ (Hg1201)\cite{kra94} and
(Cu,C)Ba$_2$Ca$_3$Cu$_4$O$_x$\cite{had97}. In particular, many
studies have been carried out on the 340 cm$^{-1}$ B$_{1g}$ phonon
in Y123 to ascertain the nature of the phonon anomaly. At this
time, however, the reason for the sensitivity of the B$_{1g}$
phonon anomaly to extremely small changes in the oxygen content
near optimal doping remains somewhat controversial
\cite{lit92,lit94}. Additionally, the physical basis of the
relationship between hole concentration and the degree of phonon
renormalization is unclear.

Subsequent to the early experiments \cite{mac87,fri90} it
was found \cite{alt91,alt91a,mcc92} that the
strength of the renormalization is very sensitive to the presence of small
amounts of impurities, even in crystals for which the critical temperature
remained close to the maximum value of 93.5 K.  For example the anomaly is very weak
in samples containing a small percentage of either Thorium \cite{alt91,alt91a}
or Gold \cite{mcc92}, where Th substitutes for Yttrium, and Au for Cu(1).
To explain this effect it was initially
suggested that the presence of impurities led to the averaging of an anisotropic
gap \cite{tho91}.

It was also known \cite{kra88} that the strength of the B$_{1g}$
phonon anomaly is very weak in samples of Y-123 with a reduced
oxygen concentration (y $<$ 6.90), and, in particular \cite{alt93}
the strength of the anomaly is very sensitive to oxygen content
near optimal doping.  For example it is very weak in crystals with
y $<$ 6.90 and T$_{c}$ = 92K, weak in crystals with y = 6.95 and
T$_{c}$ = 93.7 K, and yet very strong in overdoped crystals with y
= 7.0 and T$_{c}$ = 89.5 K.  In view of the fact that small
changes in oxygen doping produced the same effects as small
changes in impurity concentrations it was suggested
\cite{alt91a,alt93} that the strength of the B$_{1g}$ anomaly is
determined by the free carrier, or hole concentration, in the
CuO$_{2}$ planes. This suggestion was supported by the observation
\cite{che93} that the frequency of the pair-breaking peak in the
B$_{1g}$ electronic Raman continuum is very sensitive to the level
of oxygen doping and, furthermore, that its behavior could be
correlated with the strength of the phonon anomaly. That is, the
variation of the frequency of the B$_{1g}$ pair breaking peak with
doping, moved in complete step with the value of the gap obtained
from an analysis of the phonon anomaly. The dependence on hole
concentration has been revisited in a recent examination of Ca -
doped Y-123 \cite{qui00}.

In an attempt to gain additional insight into the above question,
and into the origin of the physical processes that determine the
sensitivity of the B$_{1g}$ phonon anomaly to doping, we have
carried out Raman scattering investigations of Ca-doped Y-123
(Y$_{1-x}$Ca$_{x}$Ba$_{2}$Cu$_{3}$O$_{y}$ or Y(Ca)-123).  Calcium,
a divalent alkaline-earth ion, substitutes preferentially for
trivalent Yttrium in the YBa$_{2}$Cu$_{3}$O$_{y}$ compound. Since
the ionic radius of the Ca$^{2+}$ ion is approximately equal to
that of the Y$^{3+}$ ion \cite{gre89} one expects that the carrier
or hole concentration could be varied in a controlled manner
without introducing any significant distortion in the Y-123
lattice \cite{fis93}.  Also, given that the Y-site is located
midway between the superconducting CuO$_{2}$ planes, Ca
substitution should be an effective means of increasing the
carrier concentration on the CuO$_{2}$ planes. On the basis of
simple valence considerations one might thus expect that each
substituted Ca ion would contribute 0.5 holes to each CuO$_{2}$
plane \cite{fis93,gun95}.  Although this recipe appears to break
down in the case of more heavily doped samples \cite{kuc95} (x $>$
0.1), the results presented here, which were obtained using
relatively lightly doped (x = 0.02), high-quality single crystals
of Y(Ca)-123, appear to be in accord with these expectations. Our
results indicate that the effect of calcium doping on the B$_{1g}$
phonon anomaly is equivalent in every way to the changes induced
by appropriate variation of the oxygen concentration. That is, the
magnitude of the renormalization is directly related to the hole
concentration, in contrast to recent observations \cite{qui00}. In
addition, it it found that the superconductivity induced (SCI)
frequency renormalization is small for $p < 0.15$, increases
rapidly just above optimum doping, and then increases
monotonically with increasing hole concentration (for $p > 0.15$).
The SCI linewidth renormalization changes from a narrowing below
optimal to a broadening above.  These results are consistent with
the absence of a SCI electronic renormalization in the underdoped
state where a pseudogap opens above T$_{c}$.

\section{Sample Preparation \& Characterization}
\label{PC-Y123}

Good quality Y$_{1-x}$Ca$_{x}$Ba$_{2}$Cu$_{3}$O$_{y}$ (Y(Ca)-123)
crystals were grown in yttrium-stabilized zirconia crucibles
(ZrO$_{2}$-Y) by a standard flux method \cite{lia92}.  Based on
stoichimetry considerations the [Ca] was estimated to be 4\%, but
Inductively-Coupled Mass Spectrometry (ICPMS) analysis \cite{als}
yielded a lower value of 2.0 $\pm$ 0.2 \%.

An estimate of the oxygen content (y$_{est}$) was obtained using
the crystal growth parameters of annealing pressure and
temperature \cite{sch91}.  To determine the values of $y$ more
accurately, the c-axis lattice parameter was carefully measured by
X-ray diffraction (XRD) studies. Using a Siemens D-5000
diffractometer with Cu-K$_{\alpha}$ radiation,  XRD patterns were
obtained using scans with a step size $\Delta\theta$ = 0.02
$^{\circ }$, in the range $5^{\circ } \le 2\theta \le 100^{\circ
}$. In order to obtain reliable estimates of the c-axis lattice
parameter only the $(00l)$ peaks with 2$\theta$ $>$ 30 $^{\circ }$
({\it l} $>$ 5) were used in a nonlinear least-squares fit to the
diffraction peaks. The refinement program corrects for off-axis
shifts and consequently the c-axis spacings reported in Table
~\ref{tab:parameters} are accurate to within $\pm$ 0.002 \AA.

In agreement with expectations \cite{fis93}, as mentioned above,
it was found that the c-axis lattice parameter is not affected by
small concentrations of Ca. Therefore, using the relation between
the c-axis length and the oxygen concentration \cite{alt93}, more
accurate values for the oxygen concentrations (y$_{ref}$) were
obtained (see Table ~\ref{tab:parameters}). The samples of 2 \%
Ca-doped Y123 are therefore labeled according to their oxygen
concentration (y =7.00(A), 6.93(B), 6.88(C), 6.76(D)), and an
undoped sample (y= 6.93 (U)) is also used for comparison purposes.
The critical temperature, T$_{c}$, was obtained from DC
magnetization measurements as described elsewhere \cite{lia92}.

The relationship between the oxygen concentration and hole carrier
content (p) in the crystals was deduced using a modified form of a
relation proposed by Tallon
 \cite{tal95}. He proposed that p = 0.187 - 0.21(7-y), where
p is the number of holes per CuO$_2$ layer and y is the oxygen concentration.
In this work, however, we will use a slightly
displaced value of the p-intercept to be consistent with an optimum hole
concentration of 0.16, which is obtained when y = 6.93 and T$_{c,max}$ = 93.7K.
Accordingly,
\begin{equation}
p = 0.175 - 0.21(7-y)         \label{eq:pvsy}
\end{equation}

\begin{figure}[p!]
\includegraphics{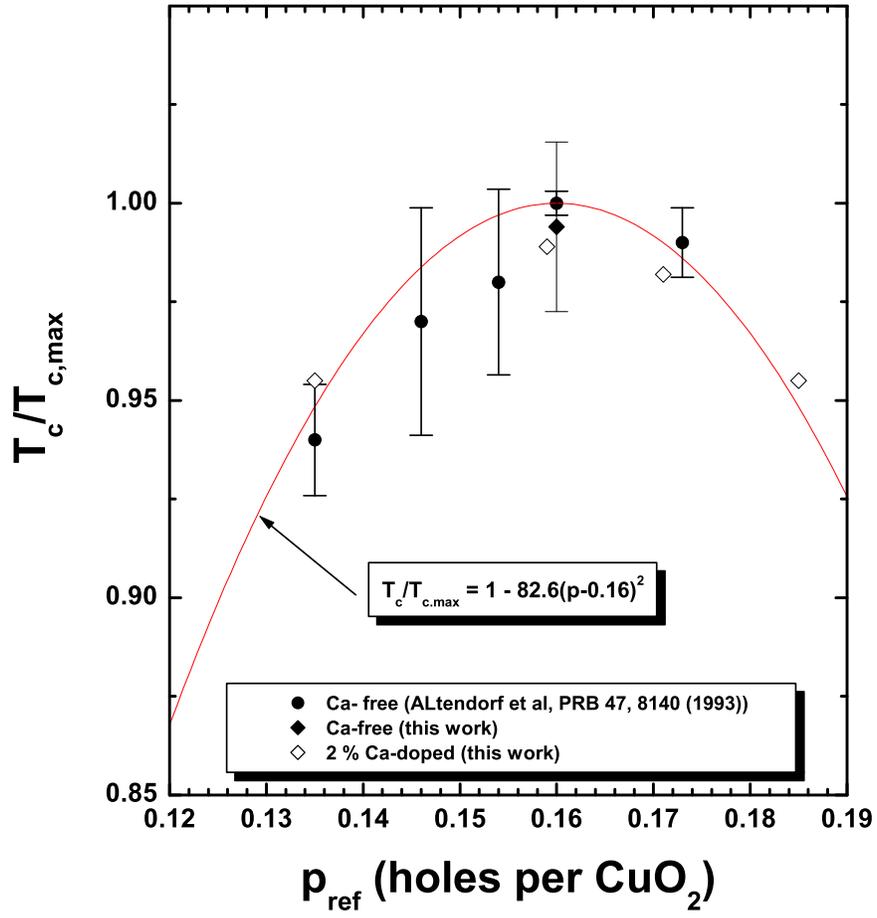}
\vspace{0.1in} \caption{The superconducting transistion
temperature versus hole concentration for Ca-doped (2\%) and
Ca-free crystals of Y123.} \label{tc-versus-p}
\end{figure}

The critical temperatures of the samples, as a function of the
hole concentration per CuO$_{2}$ (p), follow the parabolic
dependence (Fig.~\ref{tc-versus-p})reported in other
papers\cite{kuc95,tal95},
\begin{equation}
T_{c}/T_{c,max} = 1 - 82.6 (p - p_{o})^{2}     \label{eq:Tcvsp}
\end{equation}
where $p_o = 0.16$ is the optimum hole concentration at T$_{c,max}$.

It can be seen from Fig.~\ref{tc-versus-p} that crystal C is
optimally doped, D is underdoped and A and B are overdoped. We can
conclude that 2$\%$ Ca-doping of the Y-123 crystals does not
involve any noticeable oxygen depletion, which has been reported
\cite{gre89,man89,gle90,kuc95} to occur for Ca concentrations
greater than 10$\%$.  Consequently, the substitution of Calcium
(+2) for Yttrium (+3) effectively increases the hole concentration
as p = p$^*$ + [Ca]/2, where p$^*$ would be the hole concentration
in Ca-free material.  As one can conclude from an inspection of
Table ~\ref{tab:parameters}, only sample A exhibits y$_{est}$ $<$
y$_{ref}$, which may suggest that the sample is oxygen depleted.
However, since sample D is the only underdoped crystal, further
investigations are necessary to validate this conclusion.

\section{Raman spectra}

Raman spectra of the Y(Ca)-123 crystals were collected using
either the 514.5 nm or 488.0 nm lines of an argon-ion laser as the
excitation source.  To minimize local heating effects, the
incident power was kept below 3 mW, and focussed on the sample
with a spherical-cylindrical lens combination to yield incident
power densities of the order 10W/cm$^{2}$.  With this incident
power level the local sample heating is minimal. This is clear
from the fact that the observed renormalizations occur very close
to the measured critical temperatures. Spectra were obtained at
various temperatures in the range 15K$<$T$<$300K, using a Displex
refrigerator. All the sample temperatures cited in this paper are
the measured ambient temperatures.

Within the D$_{4h}$ point group, excitations of B$_{1g}$ symmetry
are selected by using, in Porto's notation, the z(x'y')z
scattering geometry, where x' denotes the (1,1,0) direction, y'
denotes the (-1,1,0) direction and z is the direction parallel to
the c-axis.  In this geometry the (1,0,0) and (0,1,0) axes lie
along the Cu-O bonds.  The scattered light was analyzed with a
triple-grating spectrometer, using gratings with a bandpass of
either 700 cm$^{-1}$ or 1400 cm$^{-1}$. The corresponding
resolution at the detector is 0.8 or 1.6 cm$^{-1}$, respectively.
The scattered light is detected by an ITT-Mepsicron imaging
detector over a typical collection time of 1 hour. The spectra
were carefully calibrated against the laser plasma lines.

\begin{figure}[ht!b!p!]
\includegraphics{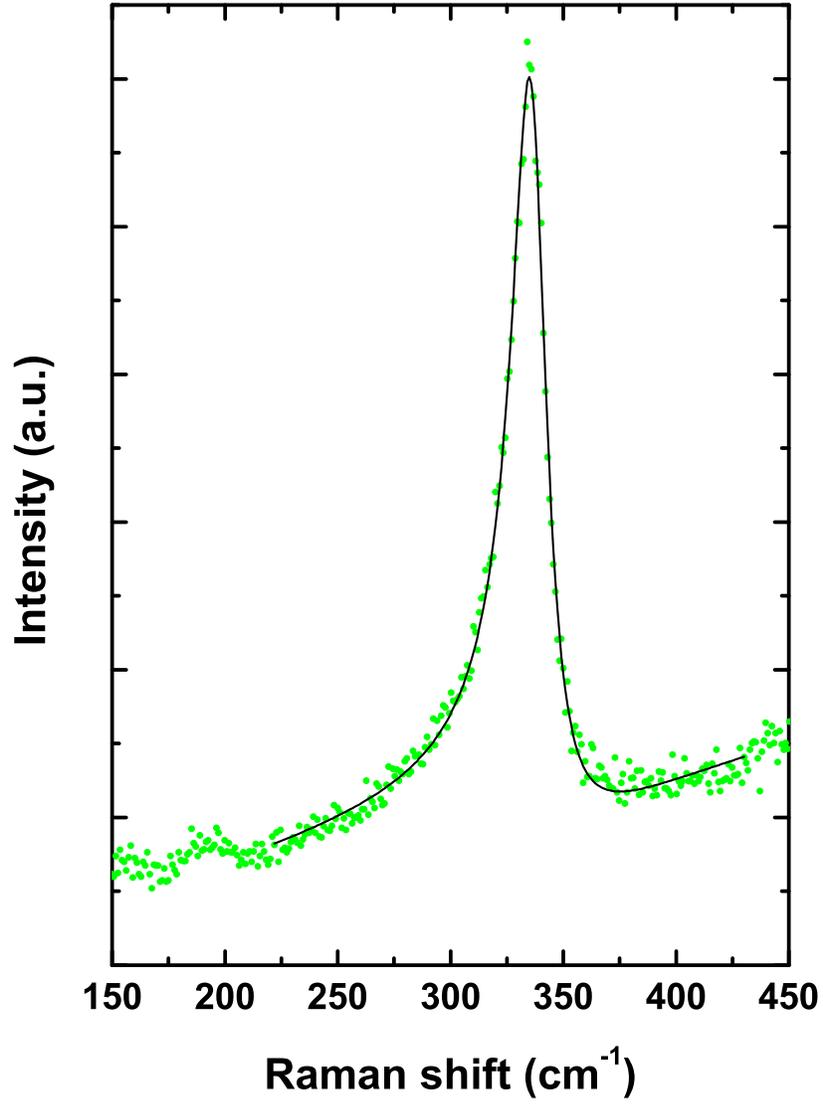}
\vspace{0.1in} \caption{Fano lineshape fit (solid line) to the
(nominal) 340 cm$^{-1}$ phonon found in the B$_{1g}$ spectra
(dots) of sample A at T = 16.6 K, resulting in $\omega = 336.3 \pm
0.1$ cm$^{-1}$, $q = -6.0 \pm 0.2$ and HWHM $\gamma = 9.6 \pm 0.1$
cm$^{-1}$.} \label{fano-A}
\end{figure}

B$_{1g}$ spectra were obtained at several different temperatures
for all samples.  Spectra at 16.6 K from sample A is shown in
Fig.~\ref{fano-A}. Quantitative analysis of the linewidth
($2\gamma = \Gamma$) and frequency ($\omega_o$) of this mode was
carried out by fitting the lineshapes to Fano profiles
\cite{gin89a} with a linear plus constant background ($b \omega +
c$):

\begin{equation}
I(\omega ) = I_o \frac{(q+\epsilon )^2}{1 + \epsilon^2} + b \omega
+ c \label{fano}
\end{equation}
where,
\begin{eqnarray}
I_o & =& \pi \varrho_o T_{e}^{2} \\ \epsilon &=& \frac{(\omega
-\omega_o)}{\gamma } \\ \gamma &=& \pi \varrho_o V^2\\ q &=&
\frac{T_p}{\pi \varrho_o V T_e}
\end{eqnarray}
Here we assume a constant density of electronic states over the
energy range of interest ($\varrho_o$), V measures the interaction
between phonon and electronic continuum states, and T$_e$, T$_p$
are the matrix elements characterizing Raman-active transitions to
the electronic continuum and phonon states, respectively. $\gamma$
is the half-width-at-half-maximum (HWHM) of the lineshape and q is
the asymmetry parameter which determines the form of the
lineshape, while b and c are adjustable constants. An example of
the results of such a fit is shown in Fig.~\ref{fano-A}.

\begin{figure}[ht!b!p!]
\includegraphics{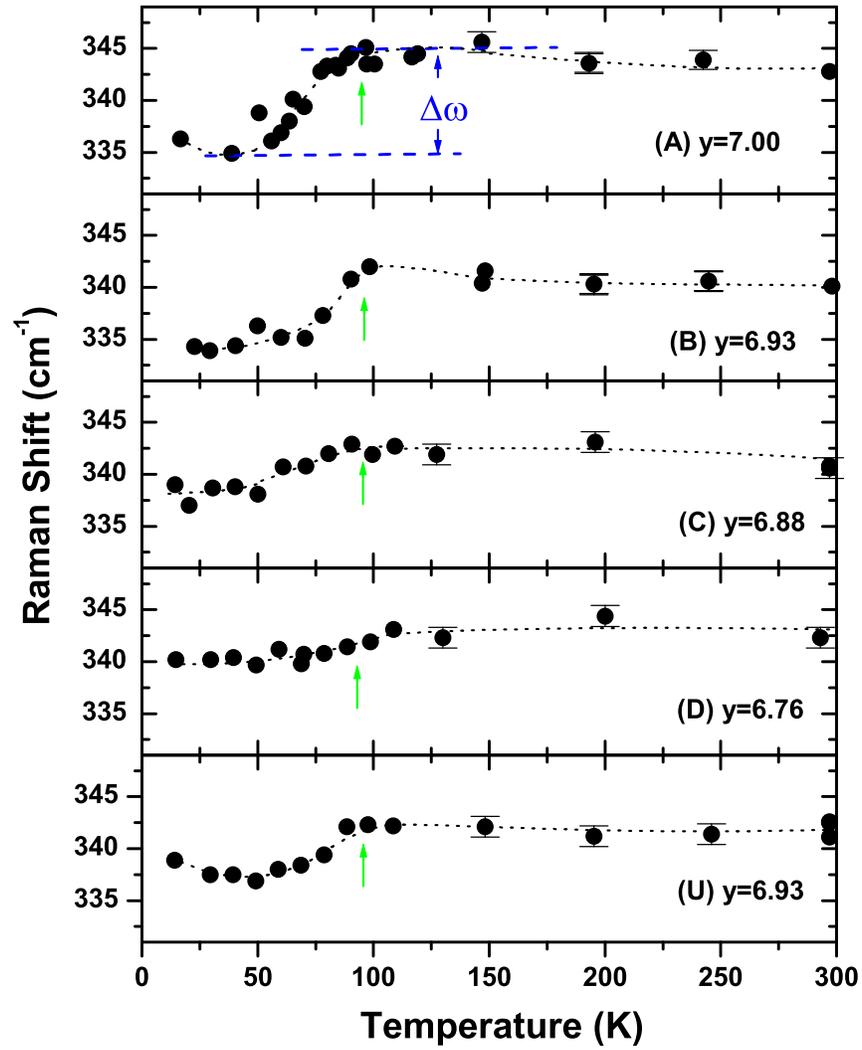}
\vspace{0.1in} \caption{The temperature dependence of the 340
cm$^{-1}$ phonon frequency, for 2 \% Ca-doped (A-D) and Ca-free
(U) crystals with varying oxygen concentrations.}
\label{om-versus-t}
\end{figure}

\begin{figure}[ht!b!p!]
\includegraphics{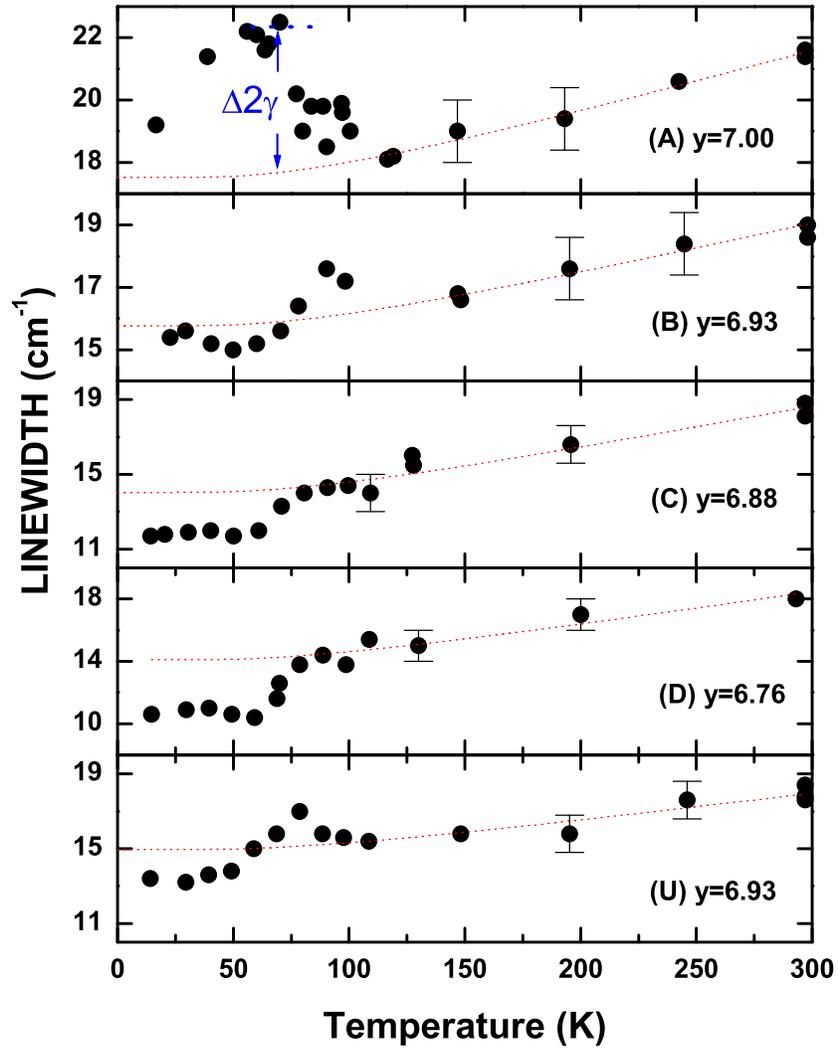}
\vspace{0.1in} \caption{The temperature dependence of the 340
cm$^{-1}$ B$_{1g}$ phonon linewidth for 2 \% Ca-doped (A-D) and
Ca-free (U) crystals with the indicated oxygen concentrations.}
\label{gm-versus-t}
\end{figure}

The temperature dependence of the frequency ($\omega_o$) and
linewidth ($2\gamma$ = FWHM) of the 340 cm$^{-1}$ B$_{1g}$ Raman
mode are summarized in Figs.~\ref{om-versus-t} and
\ref{gm-versus-t}, respectively.  As can be seen from these
figures, the (nominal) 340 cm$^{-1}$ B$_{1g}$ mode clearly shows
significant changes in linewidth and frequency as a function of
temperature.

As seen in Fig.~\ref{om-versus-t}, the magnitude of the phonon
{\it frequency} decreases substantially in crystals with lower
oxygen content. The renormalization is largest \cite{alt-note} in
the highly oxygenated samples and almost vanishes in the sample
with an oxygen content y = 6.76. There is also a substantial
softening of the phonon below 100 K, by 11 cm$^{-1}$ in crystals
with the highest oxygenation.  This softening approximately scales
with the oxygen content and in crystals with y = 6.76 the
softening is reduced to about 2-3 cm$^{-1}$.

The {\it linewidth} of the 340 cm$^{-1}$ peak
(Fig.~\ref{gm-versus-t}) decreases when cooled between room
temperature and 10 K. To determine the superconductivity induced
changes one must subtract anharmonic effects. To obtain a rough
estimate of the anharmonic changes that occur one can assume that
the phonon decays into two phonons with opposite $q$ vector, each
having a frequency $\omega_o/2$. The temperature dependence of the
linewidth can then be described approximately by the anharmonic
decay equation \cite{kle66,kle60,fri90},
\begin{equation}
\Gamma_{AH}(\omega_o,T) = c[1+2n(\omega_o/2,T)] + d
\label{eq:anharm}
\end{equation}
where $n$ is the Bose-Einstein factor, c and d are constants, and
$\omega_o$ is the frequency of the mode.  Since the Bose factor is
relatively constant for temperatures below 100K (for $\omega_o >
300 cm^{-1}$), anharmonic effects should be negligible for the 340
cm$^{-1}$ phonon in the range 100K $ > T > $ 15 K. The constants c
and d can thus be determined by fitting to the linewidth changes
that occur for T $>$ 100K, and the resulting equation can be used
to predict the anharmonic behaviour that occurs below 100K. The
superconductivity induced changes are then assumed to be the
actual linewidth minus that predicted by the anharmonic equation.
These changes are strongly dependent on the oxygen content.  In
the crystals with very high oxygenation (y = 6.98-7.00), there is
a pronounced broadening (4 cm$^{-1}$  for sample A) of the peak.
In samples (C \& D) with the lowest oxygen concentration (y = 6.88
and 6.76), a 2-3 cm$^{-1}$ step-like narrowing is observed below
80 K without any initial broadening.  In the crystal (B)
possessing an intermediate oxygen content (y = 6.93), after an
initial broadening (~1 cm$^{-1}$) below 100 K, a narrowing takes
place below 50 K (~0.5 cm$^{-1}$).

In order to quantify the SCI changes in frequency and linewidth
and to facilitate comparison with the results \cite{alt92}
obtained from Ca-free crystals, the procedures used in Ref.
\cite{alt92,fri90} will be applied to the results shown in Figs.
\ref{om-versus-t} and \ref{gm-versus-t}. That is, the magnitude of
the frequency anomaly is estimated by finding the difference in
the frequency of the phonon at two temperatures. Quantitatively,
for a given doping level, the magnitude of the phonon frequency
anomaly ($\Delta \omega$) is determined by the difference in the
phonon frequency at 30 K and 100 K,
\[
\Delta\omega \equiv \omega(30 K)-\omega(100 K),
\]
Fig.~\ref{gm-versus-t} shows that as the sample is cooled below
T$_c$ the linewidth departs from the anharmonic decay curve. The
phonon linewidth anomaly ($\Delta 2\gamma$) is measured at the
temperature T$_o$ ($< T_{c}$) where the deviation reaches its
maximum. It's magnitude is defined as the difference between the
linewidth at T$_{o}$ and the value calculated from anharmonic
decay of the phonon at the same temperature,
\[
\Delta 2\gamma \equiv 2\gamma (T_o) - \Gamma_{AH} (T_o)
\]
where, $\gamma_{AH}$ is the anharmonic linewidth given by equation
\ref{eq:anharm}.

\begin{figure}[ht!b!p!]
\includegraphics{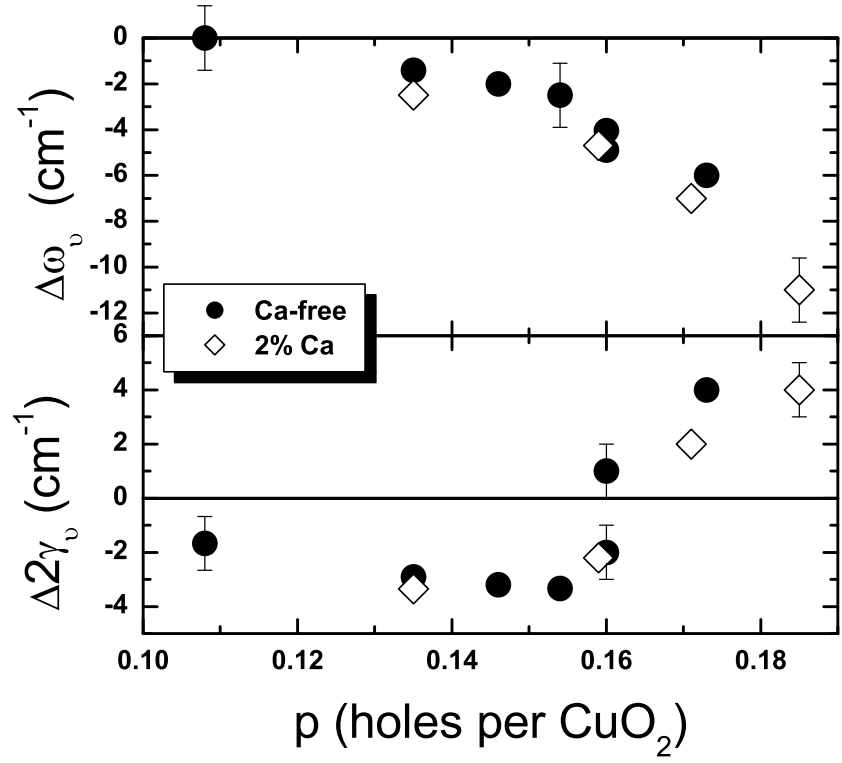}
\vspace{0.1in} \caption{Superconductivity induced renormalization
in frequency and linewidth of the 340 cm$^{-1}$ B$_{1g}$ mode, as
a function of hole concentration in Y123.} \label{wg-versus-p}
\end{figure}

When $\Delta 2\gamma$ and $\Delta \omega$ are plotted as a
function of hole concentration (Fig.~\ref{wg-versus-p}), a number
of features are evident.  First, the Ca-free and 2 \% Ca-doped
crystals fall on the same curve.  If the [Ca] were ignored, the
corresponding points in Fig.~\ref{wg-versus-p} would be shifted
(by 0.01) to lower hole concentrations, and consequently they
would not fall on the same curve as the Ca-free crystals.  This
observation allows one to conclude that it is the change in hole
concentration that is the determining factor - independent of
whether it is determined by oxygen or Ca doping.

Secondly, the most pronounced broadening ($\Delta 2\gamma = 4
cm^{-1}$) and softening ($\Delta\omega = 11 cm^{-1}$) of the 340
cm$^{-1}$ mode occurs for the sample (A) with the highest hole
concentration, estimated to be p = 0.185
(Table~\ref{tab:parameters}).  In fact, Fig.~\ref{wg-versus-p}
demonstrates that the magnitude of the frequency renormalization
monotonically increases as the hole concentration is increased.
According to theory \cite{zey88,zey90}, as the superconducting gap
energy, or the pair-breaking peak in the electronic continuum,
approaches the 340 cm$^{-1}$ phonon frequency from above, the
phonon damping and frequency renormalization should markedly
increase, as is observed.  Our results thus imply that for optimal
doping, $2\Delta
>$ $\omega_o$ = 340 cm$^{-1}$, and that the superconducting gap
decreases with increasing doping in overdoped crystals. In fact
Bock et al \cite{bock99}, in thin films with much higher Ca
concentrations, carried out measurements on samples with $2\Delta
< \omega_o$. They found that $2\Delta = \omega$ for a sample with
$p \approx  0.20$.

\section{Discussion and Conclusions}

Substituting Ca for Y in YBa$_2$Cu$_3$O$_{y}$ (6.85$\le y < $7)
has allowed us to access the overdoped regime of Y123. The
superconductivity induced renormalization of the 340 cm$^{-1}$
B$_{1g}$ phonon has been studied as a function of oxygen
concentration in both in pure and in Ca-doped crystals. In
overdoped compounds the strength of the phonon anomaly increases
as the doping level is increased above optimum (p= 0.16). This is
consistent with the known behaviour of the superconducting gap. In
optimally doped compounds the B$_{1g}$ pair breaking peak is
centered at approximately 550 cm$^{-1}$ ($2\Delta$ = 8.4kT) and
this decreases to 470 cm$^{-1}$ ($2\Delta \approx$ 7.5kT) for a
crystal with p approx 0.18 (Fig 1).  Thus the gap energy is
approaching the phonon frequency from above and the increase in
strength of the phonon anomaly is completely consistent with the
predictions of Zeyher and Zwicknagl \cite{zey88,zey90}. The
results presented here demonstrate that the behaviour in high
quality crystals containing 2$\%$ Ca is identical to that observed
in undoped samples with the same hole concentration. These
conclusions are corroborated by the results of measurements
\cite{bock99} carried out on high quality thin films of Ca-doped
Y123.  In that work it was found that $2\Delta \approx$ 340
cm$^{-1}$ for p approx 0.20. This variation in the gap energy with
hole concentration, in the overdoped regime, is very similar to
that found in Bi2212 \cite{hew01} and La214 \cite{irw99}.

As noted above, the frequency shift associated with the phonon
anomaly undergoes rather dramatic changes as the doping level
moves through optimum.  From Fig.~\ref{wg-versus-p} one can see
that for p = 0.185, a doping level slightly above optimum, $\Delta
\omega \approx $ 12 cm$^{-1}$. $\Delta \omega$ then decreases
quite rapidly to $\Delta \omega \approx$ 4 cm$^{-1}$ at optimum
(p$_o$ = 0.16). As the doping level is reduced below optimum
$\Delta \omega$ decreases more gradually, and reaches a value
($\approx$ 2 cm$^{-1}$) that is comparable to the experimental
uncertainty when p $\approx$ 0.14. The SCI linewidth change
$\Delta 2\gamma$, on the other hand, is even more dramatic in that
it abruptly changes from a positive value to a negative value
(Fig.~\ref{wg-versus-p}) as the doping level is reduced through
optimum.  These results suggest that the electron phonon
interaction decreases rapidly as the doping level is reduced
through the optimum value. This can be attributed to a
corresponding decrease in the carrier concentration, and thus the
results are consistent with investigations \cite{irw99} which show
that the pseudogap opens abruptly, and the B$_{1g}$ spectral
weight decreases dramatically, as the doping level in Y123 is
reduced below p$_o$. The absence of a phonon frequency anomaly,
and a narrowing of the linewidth, are thus completely consistent
with this picture \cite{chen97b,naei99,irw99} of the pseudogap.
The results also suggest that Ca- doping does not influence the
onset of the pseudogap, and again, in high quality samples, the
pseudogap opens at a particular hole concentration, irrespective
of how it is generated.

It is also interesting to note that the SCI renormalization of the
B$_{1g}$ electronic continuum in Y123 and La214 also vanishes
\cite{che93} rather abruptly as the doping level is reduced below
optimum, and its doping dependence has been found \cite{che93} to
be closely correlated with that of the phonon anomaly.  This is in
contrast to results obtained in Bi2212 \cite{hew01} where it is
found that a SCI renormalization of the B$_{1g}$ electronic
continuum is observed well into (p $\approx$ 0.12) the underdoped
region. This result has been attributed \cite{hew01} to an
inherent inhomogeneity in Bi2212; that is, in the underdoped
material, Bi2212 is composed of underdoped and optimally doped
regions. Given the absence of any strong variations in behaviour
of specific heat measurements \cite{luo00,tall01} near optimum
doping one might speculate that a similar inhomogeneity might be
present in Y$_{1-x}$Ca$_x$Ba$_2$Cu$_3$O$_y$ for samples with x
$\ge$ 0.10.

\section{Acknowledgements}

The financial support of the Natural Sciences and Engineering
Research Council of Canada is gratefully acknowledged.

\bibliographystyle{prsty}

\begin{table}[h!!tb]
\caption{Values of the initial oxygen concentration(y$_{est}$)
estimated from the crystal growth parameters.  The refined oxygen
concentration (y$_{ref}$) calculated from the c-axis length. p is
the hole concentrations calculated from the expression p =
0.175-0.21(7-y) and p$^{*}= p + [Ca]/2 = p + 0.01$. Crystals A,B,C
and D have $[Ca] = 0.02$, while crystal U is an undoped crystal
used for comparison purposes.}
\begin{tabular}{|c|c|c|c|c|c|c|}
 \hline
Crystal & y$_{est}$ & c $\pm  0.002$ (\AA) & y$_{ref}$ &
p$_{ref}^{*}$ & p$_{ref}$ & T$_{c}$(K) \\ \hline \hline
  A & 6.98 & 11.688 & 7.00 & 0.175 & 0.185 & 89.5  \\ \hline
  B & 6.95 & 11.698 & 6.93 & 0.161 & 0.171 & 92.0  \\ \hline
  C & 6.90 & 11.707 & 6.88 & 0.149 & 0.159 & 92.7  \\ \hline
  D & 6.85  & 11.725 & 6.76 & 0.125 & 0.135 & 89.5  \\ \hline
  U & 6.95  & 11.699 & 6.93 & 0.160 & 0.160 & 93.2  \\ \hline
  \hline
\end{tabular}
\label{tab:parameters}
\end{table}

\end{document}